\documentclass{iopart}

\usepackage{iopams}
\usepackage{graphics}
\usepackage{graphicx}
\usepackage{subfigure}
\usepackage{color}
\usepackage{epsfig}
\begin{document}

\title{A study of uranium-based multilayers: II. Magnetic properties}

\author{R Springell$^{1,2}$, S W Zochowski$^{1}$, R C C Ward$^{3}$, M R Wells$^{3}$, S D Brown$^{2,4}$, L Bouchenoire$^{2,4}$, F Wilhelm$^{2}$, S Langridge$^{5}$, W G Stirling$^{2,4}$ and G H Lander$^{6}$}

\address{$^{1}$Department of Physics and Astronomy, University College London, London WC1E 6BT, UK}
\address{$^{2}$European Synchrotron Radiation Facility, BP220, F-38043 Grenoble Cedex 09, France}
\address{$^{3}$Clarendon Laboratory, University of Oxford, Oxford OX1 3PU, UK}
\address{$^{4}$Department of Physics, University of Liverpool, Liverpool L69 7ZE, UK}
\address{$^{5}$ISIS, Rutherford Appleton Laboratory, Chilton, Oxfordshire OX11 0QX, UK}
\address{$^{6}$European Commission, JRC, Institute for Transuranium Elements, Postfach 2340, Karslruhe, D-76125, Germany}
\ead{ross.springell@esrf.fr}
\begin{abstract}
SQUID magnetometry and polarised neutron reflectivity measurements
have been employed to characterise the magnetic properties of U/Fe,
U/Co and U/Gd multilayers. The field dependence of the magnetisation
was measured at 10K in magnetic fields from -70kOe to 70kOe. A
temperature dependent study of the magnetisation evolution was
undertaken for a selection of U/Gd samples. PNR was carried out in a
field of 4.4kOe for U/Fe and U/Co samples (at room temperature) and
for U/Gd samples (at 10K). Magnetic 'dead' layers of
$\mathrm{\sim15\AA}$ were observed for U/Fe and U/Co samples,
consistent with a picture of interdiffused interfaces. A large
reduction in the magnetic moment, constant over a wide range of Gd
layer thicknesses, was found for the U/Gd system
($\mathrm{\sim4\mu_{B}}$ compared with $\mathrm{7.63\mu_{B}}$ for
the bulk metal). This could be understood on the basis of a pinning
of Gd moments arising from a column-like growth mechanism of the Gd
layers. A study of the effective anisotropy suggests that
perpendicular magnetic anisotropy could occur in multilayers
consisting of thick U and thin Gd layers. A reduction in the Curie
temperature was observed as a function of Gd layer thickness,
consistent with a finite-size scaling behaviour.
\end{abstract}

\pacs{61.12.Ha, 68.35.Ct, 68.65.Ac, 75.30.Cr, 75.30.Gw, 75.70.Cn}
\maketitle

\section{Introduction}

The past two decades have seen a new branch of condensed matter
physics develop, devoted to the investigation of multilayer thin
films, driven principally by the ability to manipulate directly the
electronic behaviour of materials. The tailoring of the multilayer
composition and the production of samples consisting predominantly
of interface regions amplify the interaction of the electronic bands
with respect to the bulk layers. The study of proximity effects at
the interfaces of these multilayer systems provides a route into the
understanding of fundamental electronic properties. Our interest
lies in the interaction of the U $\textit{5f}$ electrons with those
of the itinerant transition metal $\textit{3d}$ states of iron and
cobalt, and the more tightly bound $\textit{4f}$ electrons of
gadolinium.

An investigation of the structural and magnetic properties of the
U/Fe system has been reported previously \cite{Beesley1,Beesley2},
but recent modifications to the growth apparatus have allowed the
introduction of Nb buffer and capping layers. Paper I in this series of
articles \cite{Springell2} describes the fabrication and structural
characterisation of U/Fe, U/Co and U/Gd multilayers grown in this
way.

Earlier work on U/Fe multilayers \cite{Beesley2} showed that the
magnetism is dominated by that of the Fe atoms, and we anticipate a
similar situation in the present study. Although any polarisation of
the U atom is interesting, and is one of the objectives of this
programme of multilayer growth, we expect any such moment to be at
most $\mathrm{\sim0.2\mu_{B}}$, so that its observation requires the
use of element-specific techniques, such as X-ray magnetic circular
dichroism (XMCD) \cite{Wilhelm} or X-ray resonant magnetic
reflectivity (XRMR) \cite{Brown3}. Such techniques are not discussed
in the present paper. However, the influence of the uranium on the
growth of the multilayers, for example its atomic size relative to
the other atoms in the multilayer structure and the connection
between growth and anisotropy of the magnetic layers will be
reported. As shown in paper I, the growth of U/TM (TM = Fe, Co)
multilayers results in a poorly defined interface which has a large
amount of roughness or interdiffusion relative to the thickness of
the layer. X-ray diffraction spectra indicated that the uranium was
present in either an amorphous or in a poorly crystalline alpha-U
phase. On the other hand, in the case of U/Gd both the reflectivity
and high-angle X-ray diffraction results show that relatively sharp
interfaces are produced, and that the uranium adopts an hcp form
with a c-axis dimension not far from that of Gd, which grows in an
hcp (001) preferred orientation.

This paper will address the bulk magnetic properties and to extend
our knowledge of the structure within the U/TM and U/Gd multilayers.
A combination of magnetometry and polarised neutron reflectivity
(PNR) techniques have been used to investigate the magnetic
substructure of the ferromagnetic layers, and for the case of the
U/Gd system, to study the anisotropy and magnetic ordering (Curie)
temperature as functions of the bilayer composition.

\section{Bulk Magnetisation}

This section deals with magnetisation measurements carried out on
the U/TM and U/Gd multilayer systems. SQUID magnetometry
measurements probe only magnetisation properties from the bulk of
the sample, but it is possible to resolve effects from each of the
individual components by investigating layer thickness, t, dependent
properties. For the case of the U/Fe and U/Co samples, the
dependence of the saturation magnetisation on $\mathrm{t_{Fe}}$ and
$\mathrm{t_{Co}}$ is presented, while for the U/Gd system, a more
complete investigation has been carried out. Both $\mathrm{t_{U}}$
and $\mathrm{t_{Gd}}$ dependent studies of the ferromagnetic
saturation and of the coercive field are presented. A comparison of
the field dependent magnetisation measurements taken in the plane
and perpendicular to the plane of the film gives a measure of the
anisotropy present in these multilayers, and a zero-field cooled
(ZFC) study of the magnetisation as a function of temperature, in a
constant applied magnetic field, has been used to measure the Curie
temperature ($\mathrm{T_{C}}$) as a function of $\mathrm{t_{Gd}}$.

\subsection{Experimental method}

The magnetisation measurements were carried out in a Quantum Design
MPMS (Magnetic Property Measurement System) SQUID magnetometer both
at UCL and at the Clarendon Laboratory, Oxford. The data were taken
using both the DC and the reciprocating sample option (RSO) in
no-overshoot mode, which enables a slow, stabilisation of the field,
an important consideration when measuring ferromagnetic signals with
a large diamagnetic background, as in our case, due to the
relatively thick sapphire substrates. The measurements of the
magnetisation as a function of applied field were recorded between
-70kOe and 70kOe, at 10K, well below the Curie temperatures of iron,
cobalt and gadolinium (1043K, 1388K, and 296K respectively).

\subsection{U/TM Field dependence}

\begin{table}[htbp]\caption{\label{table1}The saturation magnetisation
values as determined by SQUID magnetometry for a selection of U/Fe
and U/Co multilayer samples. Values are included for the absolute
magnetisation, scaled by the respective areas and number of bilayer
repeats and the values of magnetic moment per (Fe,Co) atom.}
\centering
\begin{tabular}{clccr}
\br Sample Number & Composition & $\mathrm{t_{Fe, Co}}$
($\mathrm{\pm2\AA}$) & $\mathrm{M_{sat}}$ (emu/unit) &
$\mathrm{M_{sat}}$ ($\mathrm{\mu_{B}/Fe,Co}$) \\
\mr
SN71 & $\mathrm{[U_{9}/Fe_{34}]_{30}}$ & 34 & 1.09E-4 & $\mathrm{1.4\pm0.16}$ \\
SN72 & $\mathrm{[U_{23}/Fe_{17}]_{10}}$ & 17 & 8.22E-6 & $\mathrm{0.2\pm0.04}$ \\
SN74 & $\mathrm{[U_{32}/Fe_{27}]_{30}}$ & 27 & 7.86E-5 & $\mathrm{1.23\pm0.15}$ \\
SN75 & $\mathrm{[U_{35}/Fe_{27}]_{30}}$ & 27 & 7.71E-5 & $\mathrm{1.2\pm0.15}$ \\
SN76 & $\mathrm{[U_{27}/Fe_{57}]_{20}}$ & 57 & 2.32E-4 & $\mathrm{1.72\pm0.19}$ \\
SN108 & $\mathrm{[U_{28}/Co_{27}]_{20}}$ & 27 & 4.38E-5 & $\mathrm{0.94\pm0.12}$ \\
SN112 & $\mathrm{[U_{19}/Co_{19}]_{20}}$ & 19 & 9.50E-6 & $\mathrm{0.30\pm0.05}$ \\
SN114 & $\mathrm{[U_{9}/Co_{18}]_{20}}$ & 18 & 4.44E-6 & $\mathrm{0.19\pm0.05}$ \\
SN116 & $\mathrm{[U_{19}/Co_{42}]_{20}}$ & 42 & 7.87E-5 & $\mathrm{1.10\pm0.23}$ \\
\br
\end{tabular}
\end{table}

\begin{figure}[t]
\centering
\includegraphics[width=0.7\textwidth,bb=10 10 240 300,clip]{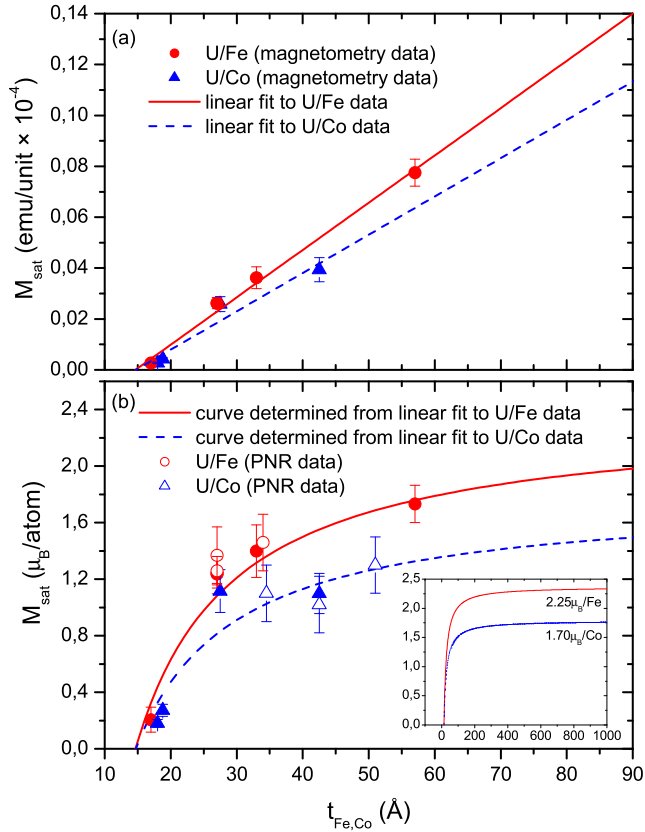}\caption
{\label{figure1}Variation of the saturation magnetisation, scaled by
the area and number of bilayer repeats, is presented in panel (a),
as a function of the ferromagnetic layer thickness. A straight-line
fit is shown for both U/Fe and U/Co systems. Values of
$\mathrm{M_{sat}}$, given in $\mathrm{\mu_{B}/Fe,Co}$, are shown in
(b) and a curve is presented, calculated from the fit in (a) and
extrapolated to large values of $\mathrm{t_{Fe,Co}}$ (insert).
Values of $\mathrm{M_{sat}}$ as determined by PNR measurements are
also presented in (b).}
\end{figure}

Table \ref{table1} lists the saturation magnetisation,
$\mathrm{M_{sat}}$ in units of emu/unit, where the values given are
normalised to the respective sample areas and number of bilayer
repeats. Values of $\mathrm{M_{sat}}$ are also given in units of
$\mathrm{\mu_{B}/Fe,Co}$. Graphs of these properties are plotted as
a function of $\mathrm{t_{Fe,Co}}$ in figure \ref{figure1}; panel
(a) describes the variation of the absolute magnetisation, while (b)
plots the values of magnetic moment per atom. A straight line fit to
the data in (a) can be made to describe the linear variation in
absolute magnetisation, which can then be scaled in the same manner
as the raw experimental data, to give the trend in saturation
moment. At large values of $\mathrm{t_{Fe,Co}}$  (insert of figure
\ref{figure1} (b)) $\mathrm{M_{S}}$ tends towards the expected bulk
moment values for Fe and Co of $\mathrm{2.2\mu_{B}}$ and
$\mathrm{1.7\mu_{B}}$ respectively. The limited number of data
points (figure \ref{figure1} (a)) implies some uncertainty in the
extremal values of $\mathrm{M_{S}}$, although it is clear that
values close to the bulk would be achieved with very thick layers.
Values of the saturation magnetisation in $\mathrm{\mu_{B}/Fe,Co}$,
as determined by polarised neutron reflectivity, are also plotted in
figure \ref{figure1} (b); these measurements are described in detail
later in the text.

The linear fit to $\mathrm{M_{sat}}$ described in figure
\ref{figure1} (a) shows a direct proportionality between the
increase in Fe or Co layer thickness and the saturation
magnetisation. The intercept of this line with the x-axis is an
indication of the thickness at which there will be no magnetic
moment, a magnetic 'dead' layer. However, this distinct separation
of a non-ferromagnetic component and a ferromagnetic one is an
unrealistic description of the Fe and Co layers. M\"{o}ssbauer
measurements on the U/Fe system \cite{Beesley2} have recently been
reinterpreted \cite{Springell2} to give an Fe layer comprising of
three components, $\mathrm{Fe_{amorphous}}$, which is paramagnetic,
$\mathrm{Fe_{bcc}}$ carrying the bulk moment, and a non-magnetic Fe
component likely to be present in a U-Fe alloy, where the spin up
and spin down $\textit{3d}$ bands of the iron can be equally
populated; this alloy will be present at both $\mathrm{U\mid Fe}$
and $\mathrm{Fe\mid U}$ interfaces, since one of the principal
factors for the formation of these alloy regions is that of chemical
interdiffusion, a process independent of the sputtering sequence.

It is not at first obvious where within the Fe layer these
components form, but careful consideration of the interfacial
structure from X-ray diffraction data \cite{Springell2} and results
from M\"{o}ssbauer spectroscopy \cite{Beesley2} suggest that the
centre of the layers are comprised of bulk-like Fe, provided the Fe
layers are thicker than the crystalline limit, $\mathrm{\sim20\AA}$
\cite{Springell2} and then an interdiffuse U-Fe alloy region will be
formed at the interfaces. These interfaces will then include
components of non-magnetic Fe, where the concentrations of Fe in U
are low, and hence the coordination number of Fe is low. Then,
across the interface from U to Fe, as the concentration of Fe atoms
increases, paramagnetic Fe will be found and then ferromagnetic Fe
as crystallites of bcc Fe form. This result is supported by the
X-ray diffraction data presented in Paper I \cite{Springell2}, which
shows that for sample SN72, $\mathrm{[U_{23}/Fe_{17}]_{10}}$, there
is no visible intensity from any bcc Fe component, however a small
magnetic moment is still present. This result suggests that the
interface is a complicated mixture of a broad spectrum of Fe
components. The similar atomic size ($\mathrm{10.3\AA^{3}}$ for Co
compared with $\mathrm{11.5\AA^{3}}$ for Fe) and diffusion
properties of cobalt are likely to result in a similar interfacial
structure and this is supported by the results shown in figure
\ref{figure1}.

Although it is not wholly accurate to assign a 'dead' layer value to
a multilayer system, it is a useful tool to compare the extent of
the interface from U/Fe to U/Co and to U/Gd. In the case of U/Fe and
U/Co the values are very similar, $\mathrm{t_{Fe}^{dead}}$ and
$\mathrm{t_{Co}^{dead}}$ are both $\mathrm{\sim15\AA}$. The
formation of such interfaces in transition metal multilayers is not
uncommon and has been reported in Ni/Cu \cite{Zheng}, Fe/W
\cite{Xiao}, Fe/V \cite{Hosoito}, Fe/Nb \cite{Mattson} and Co/Ti
\cite{VanLeeuwen} systems.

\subsection{U/Gd Field dependence}

A study of the structural properties of U/Gd multilayers
\cite{Springell2}, compared with those of U/TM systems, indicated
that the U/Gd interfaces were sharper than those for U/Fe or U/Co
multilayers and did not include such a significant region of
interdiffusion. In this case, the magnetic 'dead' layer was expected
to be small.

\begin{table}[htbp]\caption{\label{table2}Summary of the absolute
saturation magnetisation (normalised per unit area and to the number
of bilayer repeats) and $\mathrm{M_{sat}}$ ($\mathrm{\mu_{B}/Gd}$),
for all U/Gd samples measured.} \centering
\begin{tabular}{ccccc}
\br Sample Number & $\mathrm{t_{Gd}}$ ($\mathrm{\pm2\AA}$) &
$\mathrm{t_{U}}$ ($\mathrm{\pm2\AA}$) & $\mathrm{M_{sat}}$
(emu/unit) &
$\mathrm{M_{sat}}$ ($\mathrm{\mu_{B}/Gd}$) \\
\mr
SN63 & 33 & 26 & 3.67E-6 & $\mathrm{3.96\pm0.48}$  \\
SN64 & 54 & 26 & 6.04E-6 & $\mathrm{3.97\pm0.44}$  \\
SN65 & 76 & 26 & 8.63E-6 & $\mathrm{4.04\pm0.42}$  \\
SN66 & 20 & 39 & 1.74E-6 & $\mathrm{3.09\pm0.44}$  \\
SN68 & 20 & 89 & 1.46E-6 & $\mathrm{2.60\pm0.37}$  \\
SN120 & 79 & 29 & 9.32E-6 & $\mathrm{4.20\pm0.44}$  \\
SN121 & 38 & 11 & 4.23E-6 & $\mathrm{3.96\pm0.46}$  \\
SN122 & 11.4 & 11.1 & 1.03E-6 & $\mathrm{3.21\pm0.73}$  \\
SN124(600K) & 24.8 & 10.6 & 2.33E-6 & $\mathrm{3.34\pm0.46}$  \\
SN134 & 19.8 & 10 & 2.29E-6 & $\mathrm{4.11\pm0.62}$  \\
SN135 & 18.2 & 15.8 & 2.02E-6 & $\mathrm{3.95\pm0.56}$  \\
SN136 & 19.4 & 19.2 & 2.16E-6 & $\mathrm{3.96\pm0.56}$  \\
SN137 & 19.5 & 28.2 & 1.96E-6 & $\mathrm{3.57\pm0.54}$  \\
SN138 & 20 & 4.8 & 2.67E-6 & $\mathrm{4.75\pm0.67}$  \\
\br
\end{tabular}
\end{table}

\begin{figure}[htbp]
\centering
\includegraphics[width=0.7\textwidth,bb=10 10 240 300,clip]{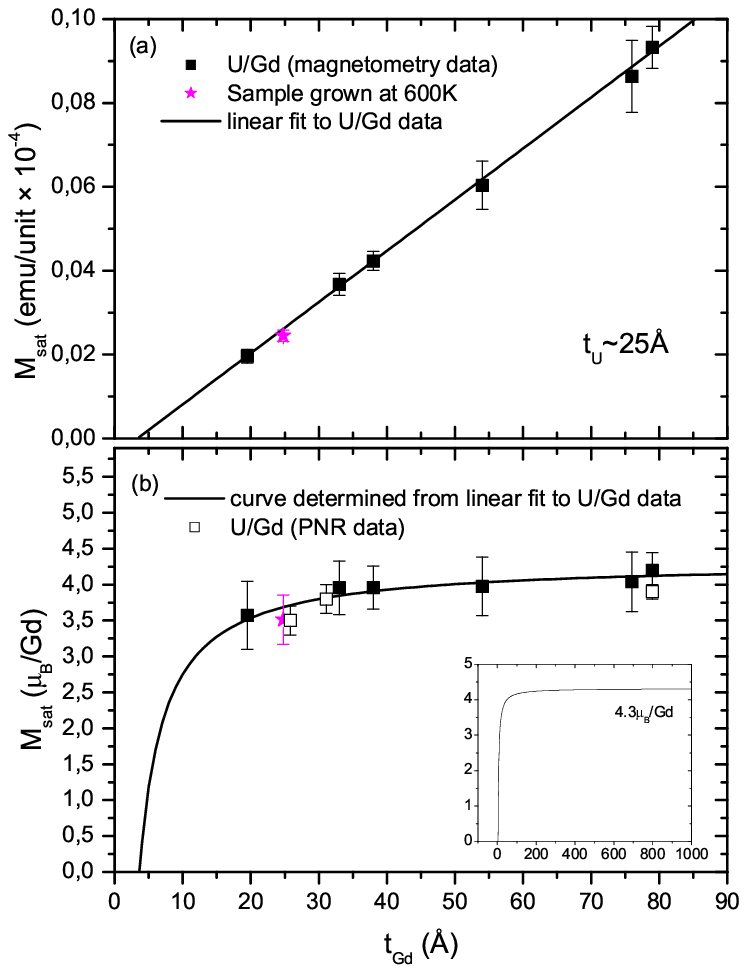}\caption
{\label{figure2}The variation of the saturation magnetisation,
scaled by the area and number of bilayer repeats, is presented in
panel (a), as a function of $\mathrm{t_{Gd}}$. A straight-line fit
is shown. Values of $\mathrm{M_{sat}}$, given in
$\mathrm{\mu_{B}/Gd}$, are shown in (b) and a curve is presented,
calculated from the fit in (a) and extrapolated to large values of
$\mathrm{t_{Gd}}$ (insert). The starred data points refer to sample
SN124, $\mathrm{[U_{10.6}/Gd_{24.8}]_{20}}$, grown at 600K. Values
of $\mathrm{M_{sat}}$ as determined by PNR measurements (open
squares) are also presented in (b).}
\end{figure}

Values of the saturation magnetisation, $\mathrm{M_{sat}}$, are
summarised in table \ref{table2}, in both units of emu/unit and
$\mathrm{\mu_{B}/Gd}$, where emu/unit refers to a normalisation of
$\mathrm{M_{sat}}$ per unit area and per bilayer. These data are
shown in figure \ref{figure2} (a) for samples with
$\mathrm{t_{U}\sim25\AA}$. A linear relationship is found between
the saturation magnetisation and the gadolinium layer thickness. The
x-axis intercept of the fitted line gives the thickness of the
magnetic 'dead' layer, which in this case is $\mathrm{\sim4\AA}$.
Figure \ref{figure2} (b) shows the variation of $\mathrm{M_{sat}}$
as a function of $\mathrm{t_{Gd}}$ in units of
$\mathrm{\mu_{B}/Gd}$. The straight-line fit to the data in panel
(a) has been scaled in the same manner as the experimental data. The
inset figure in panel (b) shows the extrapolated form of the
saturation magnetisation for large values of $\mathrm{t_{Gd}}$,
towards a value of $\mathrm{4.3\mu_{B}}$. Figure \ref{figure2} (b)
also includes saturation magnetisation data obtained from polarised
neutron reflectivity data (PNR), described in Section 3.

There are some important observations to be made from figure
\ref{figure2}. First, the magnetic 'dead' layer of about
$\mathrm{4\AA}$ is significantly thinner than that observed in the
U/TM systems. Second, whereas for the U/TM multilayers
$\mathrm{M_{sat}}$ tended towards the bulk moment value for very
thick TM layers (see insert of figure \ref{figure1} (b)) the
saturation magnetisation extrapolated for large values of
$\mathrm{t_{Gd}}$ in U/Gd multilayers tends towards a value of
$\mathrm{4.3\mu_{B}}$, significantly reduced from the bulk metal
value of $\mathrm{7.63\mu_{B}}$. This indicates that the reduction
in magnetic moment is not confined simply to the interface regions,
but is an effect arising from the bulk of the Gd layers.
Significantly reduced values of the ordered gadolinium moment have
been observed in other multilayer systems, such as Gd/Mo
\cite{Harkins}, Gd/V \cite{Pankowski} and Gd/Cr \cite{Mergia}. The
possible mechanisms for this phenomenon were discussed for the Gd/Mo
system \cite{Harkins} in terms of the Gd growth and pinning of the
Gd moments at the interface.

\begin{figure}[h!]
\centering
\includegraphics[width=0.7\textwidth,bb=10 10 225 170,clip]{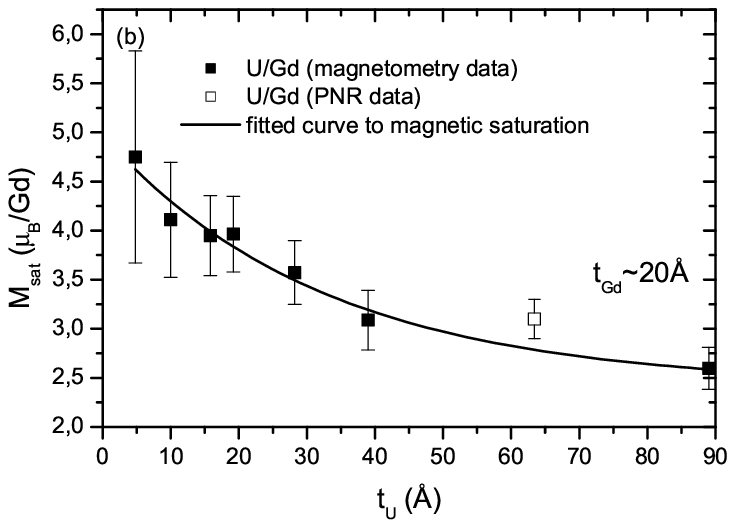}\caption
{\label{figure3}Values of $\mathrm{M_{sat}}$ are presented as a
function of $\mathrm{t_{U}}$ in units of $\mathrm{\mu_{B}/Gd}$ for
samples with $\mathrm{t_{Gd}\sim20\AA}$. A PNR measurement is shown
as the open square. The solid line is a guide to the eye.}
\end{figure}

In order to probe further the effect of the Gd structure on
$\mathrm{M_{sat}}$, a thin sputtered film of Gd was grown (sample
SN62, $\mathrm{\sim500\AA}$ thick) under the same conditions of
temperature, Ar pressure and sputtering power as the U/Gd
multilayers. A value of only $\mathrm{\sim5\mu_{B}}$ was determined;
this indicates that the majority of the magnetic moment reduction
observed in the multilayers arises from the growth mechanism of the
gadolinium layers and is not confined to the interface region.
Recalling discussions presented in paper I of this series
\cite{Springell2}, for thick Gd layers large roughnesses were
observed; these were described as step-like, a feature related to a
columnar growth modulation. It is possible that it is at the
boundaries of these column structures that the majority of the Gd
moments are pinned, although some pinning will also be present at
the interfaces.

It is clear, however, that the presence of the uranium layers
affects the magnetisation of the gadolinium layers. In order to
highlight this point figure \ref{figure3} shows the saturation
magnetisation values ($\mathrm{\mu_{B}/Gd}$) for a number of U/Gd
samples with a constant gadolinium layer thickness of
$\mathrm{\sim20\AA}$. As the U layers become thicker, so the
saturation magnetisation decays.

A likely source of this effect is again a structural one, where
variations in $\mathrm{t_{U}}$ can alter the internal strain within
the Gd layers, which can in turn affect the pinning of the Gd
moments near the U/Gd interface. A similar effect was observed in
the Gd/Mo system \cite{Harkins2}, but in this case a linear
relationship was observed between the Mo layer thickness and the
reduction in Gd moment. However, the range of Mo layer thicknesses
discussed was only between 7 and $\mathrm{15\AA}$, whereas in our
case the U spacer layer thickness varies between 5 and
$\mathrm{90\AA}$.

\subsection{Anisotropy}

The magnetic anisotropy is a measure of the preferred direction of
the magnetisation \cite{Bland}, and can be determined directly from
bulk magnetisation measurements, where the magnetic field is applied
both perpendicular to and in the plane of the film. In this case a
selected number of U/Gd samples were cut so that the measurements
could be made with the samples in both orientations. The resulting
hysteresis loops give values for the coercive fields, termed
$\mathrm{H_{C}^{para}}$ and $\mathrm{H_{C}^{perp}}$ for the two
magnetic field directions, respectively. A discussion of the
anisotropy within these systems is allied with a consideration of
the values for the coercive field, since the energy involved in
directing the moments along an easy axis is directly related to the
applied field required to rotate them. However, it should be noted
that the coercive field is also dependent on a number of other
complicating factors, such as domain size and strains within the
layers.

The magnetic anisotropy energy, $\mathrm{K_{eff}}$ can be determined
from the difference in area of the magnetisation loops; this
relationship holds for all directions of the magnetic easy axis. The
sign convention adopted for $\mathrm{K_{eff}}$, denotes that a
positive $\mathrm{K_{eff}}$ indicates a perpendicularly oriented
magnetic easy axis. This implies that the difference in area of the
two loops must be taken as the subtraction of the parallel from the
perpendicular loop.

The magnetic anisotropy energy or effective anisotropy is primarily
a combination of magnetic dipolar (favoring an in-plane alignment)
and magnetocrystalline anisotropies, stemming from the long range
magnetic dipolar and spin-orbit interactions, respectively. These
contributions can be resolved into volume, $\mathrm{K_{V}}$, and
surface, $\mathrm{K_{S}}$, contributions to give,

\begin{equation}
K_{eff}=K_{V}+2K_{S}/t_{Gd}
\end{equation}

$\mathrm{K_{V}}$ is dominated by the magnetic dipolar or shape
anisotropy, whereas the lowered symmetry at the interface leads to a
dominant spin-orbit effect and therefore surface anisotropy. A plot
of the ferromagnetic layer thickness, t, versus $\mathrm{tK_{eff}}$
generally yields a linear function, where the gradient of the line
can be used to calculate the volume anisotropy term and the
y-intercept gives twice the surface term.

Although this treatment of the anisotropy of multilayer systems has
become commonplace, we note that there are several important
assumptions concerning this extraction of the volume and surface
contributions from the effective anisotropy. First, it is assumed
that the anisotropy, localised at the interface region influences
the magnetic moments within the bulk of the layer; this is only true
if the anisotropy is much smaller than the intralayer exchange.
Also, the validity of the separation of the effective anisotropy
into surface and volume terms becomes questionable when the layers
are very thin and are almost entirely comprised of interface region.
In this case, $\mathrm{K_{V}}$ is taken as independent of the
thickness of the films, but it is possible in multilayer systems
where the lattice mismatch between the respective species is low, to
produce strain effects throughout the multilayer that introduce a
magnetoelastic contribution that changes the magnetocrystalline
anisotropy.

\begin{figure}[htbp]
\centering
\includegraphics[width=0.7\textwidth,bb=10 10 250 290,clip]{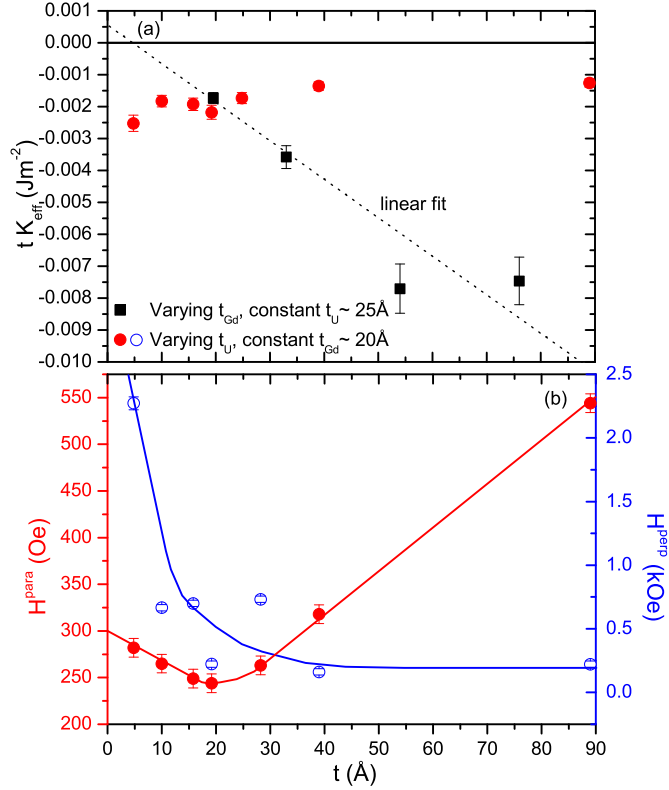}\caption
{\label{figure4}(a) $\mathrm{t_{Gd}K_{eff}}$ as a function of
$\mathrm{t_{Gd}}$ (solid black squares) and as a function of
$\mathrm{t_{U}}$ (solid red circles). The dashed black line is a
linear fit to the data. (b) coercive field values as a function of
$\mathrm{t_{U}}$ with the applied magnetic field parallel to the
plane of the sample, $\mathrm{H_{C}^{para}}$, (left-hand y-axis,
solid red circles) and perpendicular, $\mathrm{H_{C}^{perp}}$
(right-hand y-axis, open blue circles). The solid lines are guides
to the eye.}
\end{figure}

Table \ref{table3} gives $\mathrm{H_{C}^{para}}$,
$\mathrm{H_{C}^{perp}}$ and $\mathrm{K_{eff}}$ for selected U/Gd
samples. Figure \ref{figure4} (a) shows a plot of
$\mathrm{t_{Gd}K_{eff}}$ as a function of the gadolinium layer
thickness and as a function of the uranium layer thickness. Panel
(b) of figure \ref{figure4} presents the coercive field values as a
function of $\mathrm{t_{U}}$ with the applied magnetic field
parallel to the plane of the sample and perpendicular.

\begin{table}[htbp]\caption{\label{table3}Summary of the
coercive fields, determined from the hysteresis loops, taken with
the field applied parallel and perpendicular to the plane of the
film. Values of the magnetic anisotropy energy, $\mathrm{K_{eff}}$,
calculated as the difference in area of the hysteresis loops,
measured in orthogonal field directions are also listed.} \centering
\begin{tabular}{clccc}
\br Sample & Composition & $\mathrm{H_{C}^{para}}$ (Oe) &
$\mathrm{H_{C}^{perp}}$ (Oe) &
$K_{eff}$ ($\mathrm{MJ/m^{3}}$) \\
& & ($\mathrm{\pm10}$) & ($\mathrm{\pm10}$) & ($\mathrm{\pm0.01}$) \\
\mr
SN63 & $\mathrm{[U_{26}/Gd_{33}]_{20}}$ & 350 & 520 & 1.09  \\
SN64 & $\mathrm{[U_{26}/Gd_{54}]_{20}}$ & 430 & 510 & 1.43  \\
SN65 & $\mathrm{[U_{26}/Gd_{76}]_{20}}$ & 550 & 810 & 0.98  \\
SN66 & $\mathrm{[U_{39}/Gd_{20}]_{20}}$ & 320 & 160 & 0.68  \\
SN68 & $\mathrm{[U_{89}/Gd_{20}]_{20}}$ & 540 & 220 & 0.63  \\
SN134 & $\mathrm{[U_{10}/Gd_{19.8}]_{30}}$ & 270 & 670 & 0.92  \\
SN135 & $\mathrm{[U_{15.8}/Gd_{18.2}]_{30}}$ & 250 & 700 & 1.06  \\
SN136 & $\mathrm{[U_{19.2}/Gd_{19.4}]_{30}}$ & 240 & 220 & 1.12  \\
SN137 & $\mathrm{[U_{28.5}/Gd_{19.5}]_{30}}$ & 260 & 730 & 0.89  \\
SN138 & $\mathrm{[U_{4.8}/Gd_{20}]_{30}}$ & 280 & 2270 & 1.26  \\
\br
\end{tabular}
\end{table}

In bulk hcp gadolinium a small uniaxial anisotropy acts to align the
spins parallel to the c-axis between the magnetic ordering (Curie)
temperature, $\mathrm{\sim296K}$, and the spin re-orientation
transition temperature at $\mathrm{\sim230K}$. The moments then
begin to tilt away from the c-axis, reaching $\mathrm{60^{\circ}}$
at 180K and then move back towards a canting angle of
$\mathrm{30^{\circ}}$ at low temperatures \cite{Kaul}. The direction
of growth of the gadolinium layers in these U/Gd multilayers has
been well described \cite{Springell2}; the Gd layers are oriented
[001], such that the uniaxial anisotropy might act to align the
moments out of the plane of the film. Figure \ref{figure4} (a)
clearly shows that for all Gd thicknesses investigated, a negative
effective anisotropy was measured. This indicates an in-plane
alignment of the magnetic moments. These data points can be fitted
to a straight line (shown as a dashed black line in figure
\ref{figure4} (a)), which has a negative slope and a positive
y-intercept. These yield values for $\mathrm{K_{V}}$ and
$\mathrm{K_{S}}$ of $\mathrm{-1.2\pm0.3MJ/m^{3}}$ and
$\mathrm{0.28\pm0.2mJ/m^{2}}$, respectively, which are smaller than
those obtained for the U/Fe system \cite{Beesley3}. The negative
slope indicates that the dipolar magnetic anisotropy is acting to
align the moments within the plane of the film, but the positive
surface contribution suggests the presence of a magnetocrystalline
anisotropy, which acts to align the spins perpendicularly. The line
crosses the x-axis at a thickness of $\mathrm{\sim5\AA}$, thus a
$\mathrm{[U_{25}/Gd_{5}]_{N}}$ sample might exhibit perpendicular
magnetic anisotropy (PMA).

The errors on the values of $\mathrm{K_{V}}$ and $\mathrm{K_{S}}$
were relatively large and the data points at large Gd thicknesses do
not fall on the line. This may be due to roughness effects, which
are known to dramatically alter the anisotropy in multilayers. These
have been treated theoretically \cite{Bruno} and have been observed
experimentally in the Co/Pt system \cite{Kim}. A variation from
linearity at large values of $\mathrm{t_{Gd}}$ is consistent with
the observation of increased interfacial roughness for thick Gd
layers \cite{Springell2}. The effect of varying $\mathrm{t_{U}}$ on
$\mathrm{K_{eff}}$ is of a much smaller magnitude. However, for
samples with thick U layers $\mathrm{K_{eff}}$ tends towards more
positive values, favouring an alignment of the magnetic moments
perpendicular to the plane of the film.

Figure \ref{figure4} (b) shows the variation in coercive field for
different U layer thicknesses for both field directions. A minimum
in $\mathrm{H_{C}^{para}}$ and $\mathrm{H_{C}^{perp}}$ can be
observed at $\mathrm{t_{U}\sim20\AA}$. As $\mathrm{t_{U}}$ becomes
thinner $\mathrm{H_{C}^{para}}$ changes very little, whereas
$\mathrm{H_{C}^{perp}}$ shows a dramatic increase. For thicker U
layers the situation is the reverse; $\mathrm{H_{C}^{perp}}$ changes
very little, but $\mathrm{H_{C}^{para}}$ shows a marked increase.
This could be a consequence of the increased strain, a possible
explanation for the decrease in $\mathrm{M_{sat}}$ seen in figure
\ref{figure3}, which would lead to a more difficult rotation of
domains within the plane of the film.

\subsection{Temperature dependence}

The temperature dependence of the magnetisation has been measured
for a number of U/Gd multilayer samples with different gadolinium
layer thicknesses. Zero-field cooled (ZFC) measurements were made
between 5K and 375K in an applied field of 1kOe. In this way it was
possible to determine the Curie temperature ($\mathrm{T_{C}}$) of
the gadolinium layers as a function of Gd layer thickness.

It is notoriously difficult to precisely determine the ferromagnetic
transition (Curie) temperature from magnetisation measurements and
most attempts are made to describe the paramagnetic phase, using the
Curie-Weiss law for localised moments above $\mathrm{T_{C}}$. In our
case, with such a small quantity of material and a relatively large
diamagnetic background from the sapphire substrate, measurements of
the paramagnetic susceptibility at these temperatures would be very
difficult. For the purpose of this study $\mathrm{T_{C}}$ was taken
as the point at which spontaneous magnetisation begins to be
observed in the temperature dependence, where the rate of change of
the susceptibility variation is greatest, i.e. the maximum in the
second derivative of the susceptibility.

\begin{figure}[htbp]
\centering
\includegraphics[width=0.7\textwidth,bb=10 10 235 305,clip]{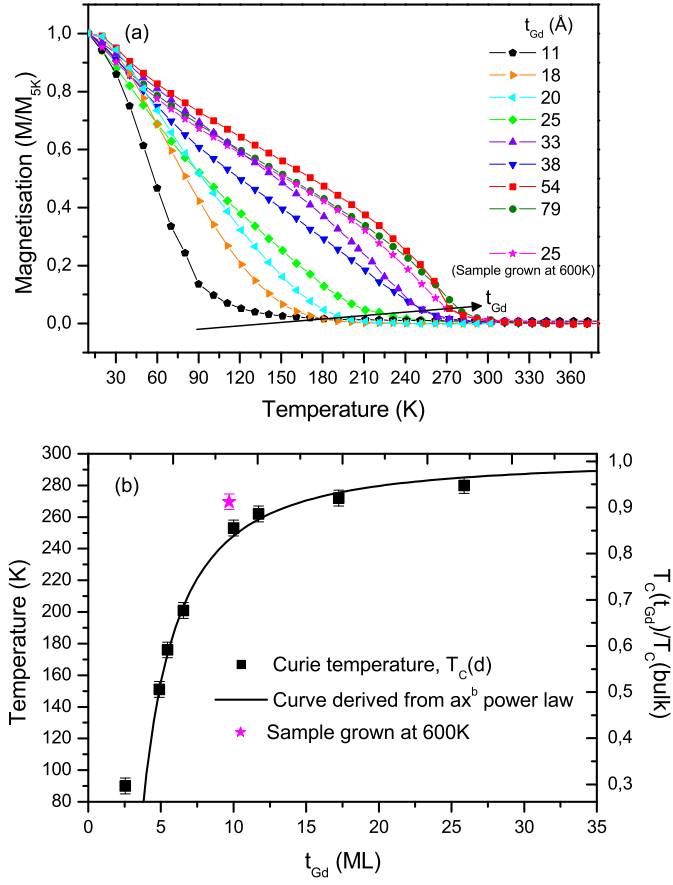}\caption
{\label{figure5}The temperature dependent evolution of the
magnetisation in an applied magnetic field of 1kOe is shown in panel
(a) for a number of U/Gd multilayer samples. The black arrow
signifies the direction of increasing Gd layer thickness and the
starred magenta data points are for the sample grown at an elevated
substrate temperature. Panel (b) plots the Curie temperature as a
function of gadolinium layer thickness. The solid black line is a
fitted curve to a finite-size scaling relationship.}
\end{figure}

Figure \ref{figure5} (a) shows the temperature dependent
magnetisation curves for a selection of U/Gd multilayers. Panel (b)
then plots the Curie temperatures for each of these samples as a
function of $\mathrm{t_{Gd}}$. The gadolinium layer thickness in
this case is presented in units of monolayers (ML), determined from
X-ray diffraction measurements \cite{Springell2} as
$\mathrm{2.9\AA}$. Such a decrease in $\mathrm{T_{C}}$ is well-known
in thin film systems and can be described by a finite-size scaling
relationship. A similar study of the thickness-dependent Curie
temperature of gadolinium grown on tungsten \cite{Farle} reported a
finite-size scaling effect and asserted that an observation of this
effect in thin films implies a layer-by-layer growth.

\begin{equation}
\frac{T_{C}(bulk)-T_{C}(t_{Gd})}{T_{C}(bulk)}=C_{0}\times
t_{Gd}^{-\lambda}
\end{equation}

$\mathrm{C_{0}}$ is an arbitrary constant, which includes
contributions from interlayer coupling effects and
$\mathrm{\lambda=1/\nu}$, where $\mathrm{\nu}$ is the
three-dimensional Ising critical exponent of the correlation length.
This treatment of finite-size scaling behaviour well describes
qualitatively, Ni \cite{Huang}, Fe \cite{Schneider}, Co \cite{Qiu}
and Gd \cite{Jiang} systems studied previously, but includes some
assumptions about the nature of the system studied. The size of the
magnetic moment is not taken into account and is assumed to carry an
equivalent value per atom for different layer thicknesses. This
requires a coherent multilayer growth with little diffusion at the
interfaces.

A fitted curve, consistent with equation (3) for the finite-size
scaling, is shown as a solid black line in figure\ref{figure5} (b)
and yields values for $\mathrm{\lambda}$ and $\mathrm{C_{0}}$ of
-1.56 and 5.75 respectively. This gives a value for the critical
exponent, $\mathrm{\nu}$, of 0.64, which is consistent with a 3D
Ising model \cite{Amazonas} and is consistent with the relationship
observed in the Gd/W thin films \cite{Farle}. These values are
quoted within an error of 5\%, due to uncertainties in
$\mathrm{\mathrm{T_{C}}}$ and $\mathrm{t_{Gd}}$. It is also noted
that the presence of capping layers on thin ferromagnetic films
\cite{Poulopoulos2} can have an appreciable effect on
$\mathrm{T_{C}}$ due to modifications in the bandstructure from
hybridisation effects at the interface.

For U/Gd multilayers with the thinnest gadolinium layers, the
observed $\mathrm{T_{C}}$ is much higher than that expected from the
finite-size scaling behaviour. This has also been observed in the
Gd/W system \cite{Farle}, \cite{Li} and has been related to the
cross-over from 3D to 2D magnetic behaviour, where theory predicts
$\mathrm{\nu=1.00}$ in the 2D regime. Sample SN124, grown at a
temperature of 600K, also exhibits a $\mathrm{T_{C}}$ that is higher
than that expected. This follows observations of the Gd/W system
\cite{Farle}, where the increased $\mathrm{T_{C}}$ values for
respective Gd layer thicknesses were attributed to the accommodation
of misfit dislocations and the presence of large inhomogeneous
strains, caused by steps and other defects at the interface.
Previous results have shown that as the substrate temperature is
increased the crystallinity is improved, but the layer roughness
increases \cite{Springell2}. This is consistent with a columnar
growth and step-like roughness profile, which could produce the
effects necessary for an elevated Curie temperature.

\section{Polarised Neutron Reflectivity (PNR)}

Neutrons are an extremely useful tool to simultaneously study the
chemical and magnetic structure, since they interact with both the
nuclei and atomic moments. Polarised neutron reflectivity can be
used to detect the spin-flip (SF, $\mathrm{R^{+-}}$ and
$\mathrm{R^{-+}}$) and non-spin-flip (NSF, $\mathrm{R^{++}}$ and
$\mathrm{R^{--}}$) scattering channels over a typical Q range of
$\mathrm{0.005-0.25\AA^{-1}}$. In this way, PNR can be used to
determine the layer thicknesses and roughnesses, the average
magnetic moment per atom, the orientation of the magnetisation and
the distribution of magnetisation in magnetic multilayers.

\subsection{Experimental method}

PNR measurements were carried out on the CRISP reflectometer of the
ISIS time-of-flight neutron source at the Rutherford Appleton
Laboratories in Chilton, Oxfordshire. The CRISP instrument is
situated after the liquid hydrogen moderator at 20K and analyses a
wavelength band of 0.5-6.5 $\mathrm{{\AA}}$. The neutron beam
arrives at the experimental hutch at an inclination angle of
$1.5^{\circ}$ and the entrance slit provides a cross-section that is
40mm wide and between 0.5 and 6mm high. The wavelength band is
defined by a disc chopper with an aperture and a nimonic chopper is
used for pulse suppression. A non-adiabatic spin flipper is used to
select the neutron spin direction and polarisation relaxation is
suppressed by a guide field to the sample position. The sample sits
approximately 10m from the moderator and reflects neutrons through
two further slits to the detector, 1.75m away.

U/Fe and U/Co samples were measured at room temperature and PNR data
were collected in the on-specular geometry in an applied magnetic
field of 4.4kOe; large enough to saturate the magnetisation within
the plane of the film. U/Gd samples were cooled (ZFC) to 10K, well
below the their respective Curie temperatures. Since the
measurements were made at magnetic saturation, only the NSF channels
were collected. The xPOLLY programme \cite{XPolly} was used to
calculate polarised neutron reflectivities for given samples and
these were fitted to the experimental data.

\subsection{Results}

The following section presents the experimental data and calculated
reflectivities for a selection of U/Fe, U/Co and U/Gd samples. The
layer thickness and roughness parameters were fixed at values
obtained from X-ray reflectometry. For the case of the U/TM
multilayers the large roughnesses accounted for the diffusion at the
interfaces, responsible for the magnetic 'dead' layer observed in
the bulk magnetisation data. In these systems the ferromagnetic
layers were separated into regions, which exhibit a magnetic moment
per atom equivalent to the bulk metal, placed at the centre of the
layers, and regions with no magnetic moment, at the interfaces.
Although it is clear that the layers will not have such discrete
sections, this technique applied to these samples was not sensitive
to more complex distributions of the magnetisation.

\begin{figure}[htbp]
\centering
\includegraphics[width=1.0\textwidth,bb=10 10 290 210,clip]{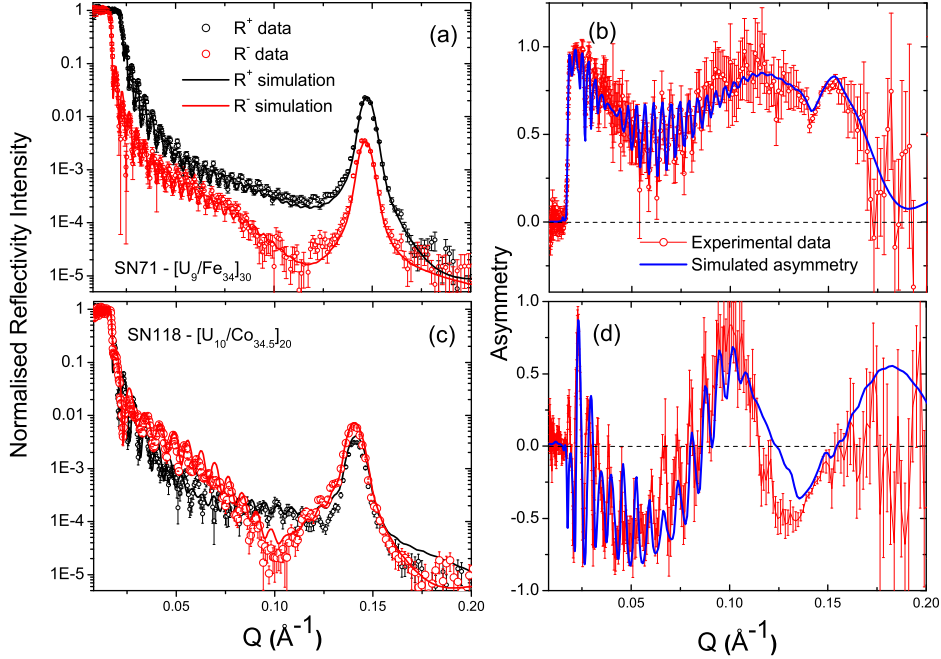}\caption{\label{figure6}
Panels (a) and (c) present the polarised neutron reflectivity data
measured in the specular geometry at 300K and 4.4kOe for samples
$\mathrm{SN71-[U_{9}/Fe_{34}]_{30}}$ and
SN118-$\mathrm{([U_{10}/Co_{34.5}]_{20})}$, respectively.
Experimental data and fitted, calculated curves are shown as black
points (curve) for the $\mathrm{R^{++}}$ channel and red points
(curve) for the $\mathrm{R^{--}}$. The asymmetry for these two
samples is shown in panels (b) and (d); red points represent the
data and the solid blue line, the fitted calculation.}
\end{figure}

It is possible to accurately determine the average magnetisation per
atom, by closely monitoring the splitting of the two spin channels
at the critical edge; the point at which the neutron transmits
through the multilayer and no longer undergoes total external
reflection. It is in this region of the reflectivity curve that the
splitting of the two spin channels is sensitive to the magnetisation
of the whole sample. The values for the average magnetic moment per
atom have been displayed in figures \ref{figure1} (b), \ref{figure2}
(b) and \ref{figure3}.

Figure \ref{figure6} (a) shows the normalised non-spin-flip
scattering from an example U/Fe sample, SN71
$\mathrm{([U_{9}/Fe_{34}]_{30})}$. The solid lines represent the
fitted calculated reflectivities for both spin channels. Panel (b)
displays the asymmetry, determined from the difference divided by
the sum of the reflected intensities. Figure \ref{figure6} (c) and
(d) present similar results for the U/Co sample, SN118
$\mathrm{([U_{10}/Co_{34.5}]_{20})}$.

\begin{table}[htbp]\caption{\label{table4}The thicknesses of the three
components of the transition metal layers for a selection of U/Fe
and U/Co samples, determined from the fitted calculations of the
PNR.}\centering
\begin{tabular}{clcccc}
\br Sample & Composition & $\mathrm{t_{TM1}}$ ($\mathrm{\AA}$)& $\mathrm{t_{TM2}}$ ($\mathrm{\AA}$)& $\mathrm{t_{TM3}}$ ($\mathrm{\AA}$)\\
\mr
SN71 & $\mathrm{[U_{9}/Fe_{34}]_{30}}$ & 8.0 & 22.5 & 3.5  \\
SN74 & $\mathrm{[U_{32}/Fe_{27}]_{30}}$ & 8.5 & 16.2 & 2.5  \\
SN75 & $\mathrm{[U_{35}/Fe_{27}]_{30}}$ & 7.5 & 15.0 & 4.5  \\
SN116 & $\mathrm{[Co_{42}/U_{19}]_{20}}$ & 11.0 & 24.0 & 7.0  \\
SN117 & $\mathrm{[U_{9}/Co_{51.3}]_{15}}$ & 8.0 & 39.0 & 4.3  \\
SN118 & $\mathrm{[U_{9}/Co_{34.5}]_{20}}$ & 8.0 & 22.5 & 4.0  \\ \br
\end{tabular}
\end{table}

A model, describing a multicomponent ferromagnetic layer in U/TM
multilayers has been proposed in an earlier investigation of U/Fe
multilayers \cite{Beesley2}, but a new interpretation of these
components was given in paper I \cite{Springell2}
($\mathrm{UTM_{alloy}|TM_{amorphous}|TM_{bulk}|TMU_{alloy}}$). Table
\ref{table4} gives the thicknesses for a three component
ferromagnetic transition metal layer. TM1 represents a combination
of the $\mathrm{UTM_{alloy}}$ and $\mathrm{TM_{amorphous}}$ regions
and carries no magnetic moment. It has a density that is
$\mathrm{5\%}$ reduced from that of the bulk metal. The central
region of the layer, TM2, has equivalent magnetisation and density
properties to the bulk. TM3 includes the $\mathrm{TMU_{alloy}}$. The
calculated reflectivities and asymmetries only well reproduce the
experimental data when TM1 is greater than TM3; a trend observed for
all U/TM samples investigated. This is consistent with the current
description of the growth of U/TM multilayers \cite{Springell2}. The
sum of the thickness of layers TM1 and TM3 give the size of the
magnetic 'dead' layer, $\mathrm{\sim12\AA}$, which is close to the
value determined by SQUID magnetometry ($\mathrm{\sim15\AA}$).

\begin{figure}[htbp]
\centering
\includegraphics[width=1.0\textwidth,bb=10 10 280 130,clip]{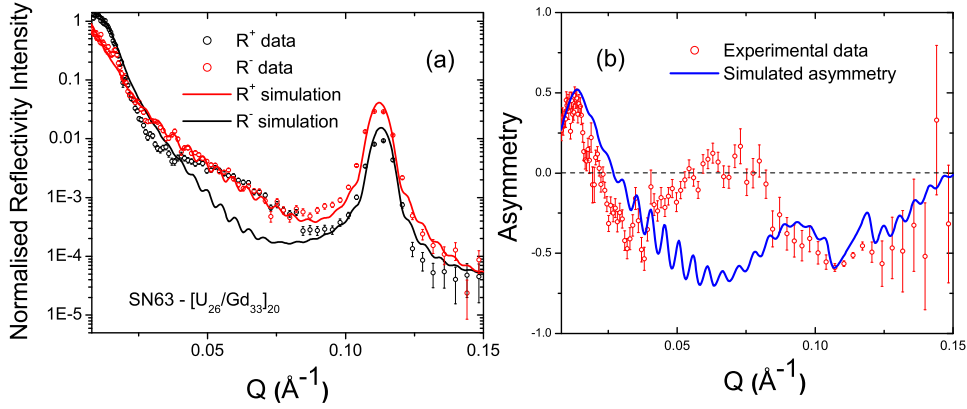}\caption{\label{figure7}
The polarised neutron reflectivity data (panel (a)) and the
asymmetry (panel (b)), measured in the specular geometry at 10K and
in an applied magnetic field of 4.4kOe are presented for sample
$\mathrm{SN63-[U_{25}/Gd_{33}]_{20}}$. Experimental data and fitted,
calculated curves are shown as black points (curve) for the
$\mathrm{R^{++}}$ channel and red points (curve) for the
$\mathrm{R^{--}}$. The asymmetry data is represented by red points
and the fitted calculation, by the solid blue line.}
\end{figure}

\begin{figure}[htbp]
\centering
\includegraphics[width=1.0\textwidth,bb=10 10 280 130,clip]{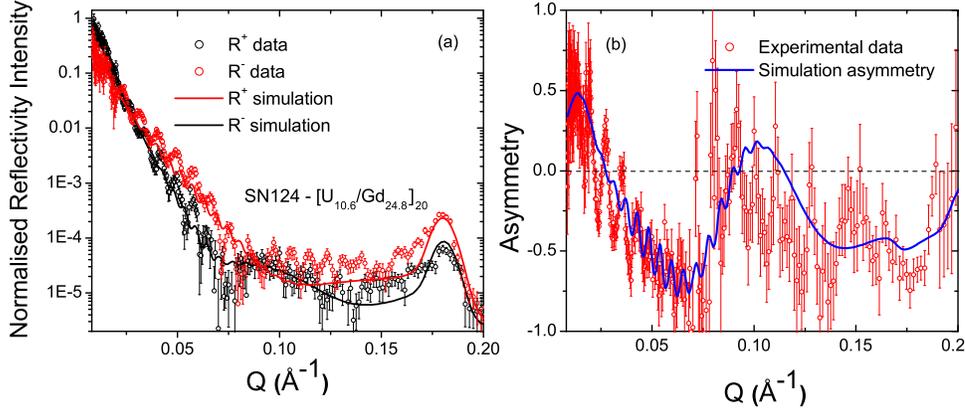}\caption{\label{figure8}
The polarised neutron reflectivity data (panel (a)) and the
asymmetry (panel (b)), measured in the specular geometry at 10K and
in an applied magnetic field of 4.4kOe are presented for sample
$\mathrm{SN124-[U_{10.5}/Gd_{24.8}]_{20}}$. Experimental data and
fitted, calculated curves are shown as black points (curve) for the
$\mathrm{R^{++}}$ channel and red points (curve) for the
$\mathrm{R^{--}}$. The asymmetry data is represented by red points
and the fitted calculation, by the solid blue line.}
\end{figure}

Figure \ref{figure7} and figure \ref{figure8} show the experimental
data and fitted calculations for the NSF scattering of samples SN63
($\mathrm{[U_{25}/Gd_{33}]_{20}}$) and SN124
($\mathrm{[U_{10.5}/Gd_{24.8}]_{20}}$). The latter was grown at an
elevated substrate temperature of 600K.

The U/Gd samples were modelled by a simple bilayer structure. Figure
2 (b) showed that the saturation magnetisation did not vary
significantly for a wide range of Gd layer thicknesses, indicating a
constant distribution of the magnetisation within the Gd layers. The
reduced values of the magnetic saturation calculated from SQUID
magnetometry measurements were supported by those obtained from the
calculated reflectivities, fitted to the PNR data. These values are
listed in table \ref{table5} for a selection of U/Gd samples.

\begin{table}[htbp]\caption{\label{table5} Summary of the input parameters
used to calculate the polarised neutron reflectivity data, fitted
using the xPOLLY \cite{XPolly} programme. Values for the respective
layer thicknesses, densities and magnetic moments are included.}
\centering
\begin{tabular}{cccccc}
\br Sample & Layer & b ($\mathrm{\times10^{-5}\AA}$) & N ($\mathrm{\times10^{28}atoms/m^{3}}$) & t ($\mathrm{\AA}$) & $\mathrm{\mu_{B}/Gd}$ \\
\mr
SN63 & Gd & 6.500 & 2.6 & 31.1 & 3.9 \\
 & U & 8.417 & 4.8 & 25.0 & 0.0 \\
SN65 & Gd & 6.500 & 3.0 & 79.0 & 3.8 \\
 & U & 8.417 & 4.6 & 26.0 & 0.0 \\
SN67 & Gd & 6.500 & 3.0 & 63.4 & 3.1 \\
 & U & 8.417 & 4.8 & 20.0 & 0.0 \\
SN124 & Gd & 6.500 & 2.7 & 24.8 & 3.9 \\
 & U & 8.417 & 4.8 & 10.0 & 0.0 \\ \br
\end{tabular}
\end{table}

The input parameters consist of the real and imaginary parts of the
neutron scattering length, b ($\mathrm{\times10^{-5}\AA}$), the
atomic number density, N ($\mathrm{\times10^{28}atoms/m^{3}}$), the
magnetic moment per atom ($\mathrm{\mu_{B}/atom}$), the angle
between the moments and the applied field, theta (taken to be
$\mathrm{0^{\circ}}$ in our case, since a magnetic field large
enough to saturate the gadolinium layers was applied), and the layer
thickness, t ($\mathrm{\AA}$).

It is clear from figures \ref{figure7} (a) and (b) that the
calculated reflectivities for the NSF channels do not well reproduce
the data far away from the Bragg peak positions. In particular, the
$\mathrm{R^{--}}$ intensity is significantly larger in this region
than that calculated. This results in a large asymmetry in the
calculated case, but very little splitting of the NSF channels in
the experimental data. The saturation magnetisations, determined
from the splitting of the channels in the vicinity of the critical
edge are consistent with SQUID magnetometry measurements. These
measurements indicated an even distribution of the magnetic moment
through the Gd layers. It is therefore likely that the discrepancy
between calculated and experimental PNR data is a product of the
mechanism for the reduced magnetic moment. Attempts to introduce a
complex interfacial structure, as was successful for the U/TM
samples, did not result in any improvement in the fit for the U/Gd
multilayers.

\section{Conclusions}

The magnetic properties of U/TM (Fe and Co) and U/Gd multilayers
have been determined, by a combination of SQUID magnetometry and
polarised neutron reflectivity measurements. These results are
consistent with ideas developed in paper I \cite{Springell2},
concerning the structure and interfacial properties of U/TM and U/Gd
systems.

Magnetisation measurements on U/Fe multilayers grown on sapphire
substrates with Nb buffer and capping layers, exhibited similar
magnetic properties to those grown on glass \cite{Beesley2},
indicating that the predominant effects in this system arise from
the interfaces of these two elements. Qualitatively similar effects
were reported on the U/Co system. For large $\mathrm{t_{Fe}}$ and $\mathrm{t_{Co}}$, values for the saturation magnetisation close to that of the bulk metals were predicted. The presence of a magnetic 'dead' layer ($\mathrm{\sim15\AA}$) for U/TM multilayers was determined by
both SQUID magnetometry and PNR. However, it was possible to
determine a further degree of complexity concerning the TM layers,
from the PNR data. The experimental data were well modelled by
slicing the TM layers into regions with no magnetic moment either
side of a central, bulk-like region. The thicknesses of these slices
were consistent with the structural description of the U/TM
interfaces developed in paper I \cite{Springell2}.

A 'missing' moment was also discovered in the U/Gd system,
determined by SQUID magnetometry measurements and PNR. However, the
mechanism for this reduction is clearly different from that observed
in U/TM multilayers. The 'dead' layer reported in this case is
$\mathrm{<4\AA}$; too small to account for such a large loss in
magnetisation. In fact, this reduction was constant across a wide
range of Gd layer thicknesses and was even observed for sputtered
thin films of gadolinium. Hence, this is not an effect confined to
the U/Gd interfaces, but one that is present throughout the whole of
the Gd layers. A column-like growth, capable of the necessary
strains and resultant defects in order to pin large numbers of Gd
moments is the most likely explanation for this reduction; a
picture, which is consistent with observations of the roughness and
crystalline properties described in paper I \cite{Springell2}.

A study of the anisotropy in U/Gd multilayers was carried out as a
function of both $\mathrm{t_{U}}$ and $\mathrm{t_{Gd}}$. A precise
description of these trends is complex, since a large number of
effects can alter the magnetic anisotropy; these effects, which
commonly relate to the thickness of the respective layers, are often
correlated to one another. However, the dependence of
$\mathrm{K_{eff}}$ upon the Gd layer thickness, clearly results in a
negative volume contribution and a small positive surface contribution
(although the error bars are large), favouring in-plane and
out-of-plane alignments of the magnetic moments, respectively. The
general trend as a function of $\mathrm{t_{U}}$ is for an out-of
plane alignment for thick U layers. Samples with thin Gd layers and
thick U present a possible route towards PMA.

A close inspection of the PNR data and calculated reflectivities for
the U/Gd system revealed a discrepancy between them at the
half-Bragg peak positions. This feature was represented by an
unusually large intensity of the $\mathrm{R^{--}}$ channel, compared
with that calculated. Explanations at this point stem from a
discussion of the mechanisms responsible for the large reduction in
saturation moment; that the pinned moments at the boundaries of
column-like gadolinium grains provide a certain amount of diffuse
scattering, which could contribute to the intensity observed in the
specular. Further PNR measurements are planned on thickness matched
U/Gd samples, in order to suppress the intensity from the even order
Bragg peaks and improve the sensitivity to the magnetic scattering
at these positions. Off-specular measurements are planned to
investigate the diffuse scattering so that a model can be developed,
which can explain such a large reduction in $\mathrm{M_{sat}}$. A
programme of measurements is also underway to investigate the
magnetism of uranium in these multilayer systems, utilising the
element-selectivity of synchrotron radiation.

\ack

RSS acknowledges the receipt of an EPSRC research studentship. We
would like to thank Keith Belcher and Peter Clack of the Clarendon
laboratory in Oxford and Tim Charlton and Rob Dalgleish of the CRISP
beamline at the ISIS neutron source.

\section*{References}
\bibliographystyle{unsrt}
\bibliography{Umultilayers2}
\end{document}